\documentclass[a4paper]{article}
\usepackage{geometry}
\usepackage{graphicx} 
\usepackage{caption}
\usepackage{subcaption}
\usepackage{booktabs}
\usepackage{array}
\usepackage{setspace}
\usepackage{appendix}
\usepackage{longtable}

\usepackage[style=apa,natbib=true]{biblatex}
\AtEveryBibitem{%
  \clearfield{issn} 
  \clearfield{note} 
  \clearfield{urldate}
  \ifentrytype{misc}{}{
    \clearfield{url}
  }
}
\title{Large Language Models Are Democracy Coders with Attitudes}
\author{Nils B.~Weidmann \and Mats Faulborn \and David Garc{\'i}a\\Department of Politics and Public Administration\\University of Konstanz, Germany}

\date{\today}

\begin{document}

\maketitle

\begin{abstract}
Current political developments worldwide illustrate that research on democratic backsliding is as important as ever. A recent exchange in Political Science \& Politics (2/2024) has highlighted again that the measurement of democracy remains a challenge. With many democracy indicators consisting of subjective assessments rather than factual observations, trends in democracy over time could be due to human biases in the coding of these indicators rather than empirical facts. In this paper, we leverage two cutting-edge Large Language Models (LLMs) for the coding of democracy indicators from the V-Dem project. With access to huge amounts of information, these models may be able to rate the many ``soft'' characteristics of regimes at substantially lower costs. While LLM-generated codings largely align with expert coders for many countries, we show that when these models deviate from human assessments, they do so in different but consistent ways: Some LLMs are too pessimistic, while others consistently overestimate the democratic quality of these countries. While the combination of the two LLM codings can alleviate this concern, we conclude that it is difficult to replace human coders with LLMs, since the extent and direction of these attitudes is not known a priori.
\end{abstract}

\onehalfspacing

The measurement of democracy has long been a contested subject of investigation in political science. A recent symposium in Political Science \& Politics (Volume 57, Issue 2) picks up on this question and discusses whether the observation of a global trend of democratic backsliding could be due to subjective perceptions of human coders. Focusing on V-Dem, the largest collection of democracy data, the argument is that expert ratings in this dataset could be affected by psychological biases \citep{little2024measuring,treisman2024psychological} and lead to skewed assessments of democratic quality that are not supported by more factual observations of institutional characteristics (see, however, \citealt{knutsen2024conceptual} for a critique).

Rather than dismissing human coding in general, in this paper we study if and how it could be supplemented with automatic coding. For political science, the advent of AI and in particular, Large Language Models (LLMs), has brought many new opportunities. The most promising and most frequent way in which these models are used is the generation of new data for empirical research, assisting (or sometimes even replacing) humans as creators or sources of these data. Much work in this area has shown that LLMs can reproduce responses collected in surveys \citep{mellon2024ais} or voting decisions by different groups of voters \citep{vonderheyde2024united}, thus reducing the need to obtain large population samples. In this paper, we study the use of LLMs for human coding in comparative research on democracy, which is a second way in which AI can assist in the creation of typically human-sourced data for social science research. While LLMs may be able to avoid some of the biases that human coders typically display, this approach could also reduce the costs of coding dramatically. The exercise we present here (the coding of one year's worth of indicators from the V-Dem dataset) was completed with a budget of less than EUR 150, while employing human coders for the same task probably cost several tens of thousands of EUR. However, the key question is whether LLMs produce codings of sufficient quality.  

We show that, perhaps not surprisingly, LLMs can emulate human coders well, and that they can do so off-the-shelf, without any adaptation. However, we also show that LLMs are of little help with those countries that V-Dem coders find particularly difficult and where they disagree the most. What is even more concerning is that while LLMs may not exhibit the cognitive biases typically attributed to humans, they have other, extremely pronounced issues: Results show that one LLM in our study consistently underestimates democratic quality, another overestimates it in almost all cases. Hence, LLMs seem to have particular political ``attitudes'' that strongly affect their coding. Since we do not know the direction and strength of these attitudes, it is difficult to trust assessments generated by particular LLMs alone or by combinations of them.

\section*{Coding Democracy is a Difficult Task}

How does democracy coding compare to other human coding tasks, and why could LLMs be particularly helpful in this task? We start with a general categorization. In human coding, coders (typically experts) are tasked with the creation of standardized \textit{codings} for particular \textit{instances/cases}. For a simple categorization of human coding tasks, we can distinguish by (i) the extent to which the material required for the coding is provided as part of the task, and (ii) the extent of interpretation that is required by the coder to perform the task. The first dimension refers to whether the coding is based on material that is readily provided to the coder. Some coding tasks involve particular instances of coding material that the coders then work with: For example, annotations of social media posts \citep{gilardi2023chatgpt} or the coding of protest events from news reports \citep[Ch. 4]{weidmann19internet} provide coders with material that the coding is supposed to be based on. Other coding tasks, such as the human coding of democracy indicators \citep{coppedge2024vdem} or the fact-checking of particular statements \citep{ni2024afactaa}, often provide no material whatsoever, and assume that coders possess expert knowledge to perform the task. 

\begin{figure}
    \centering
    \includegraphics[width=0.7\linewidth]{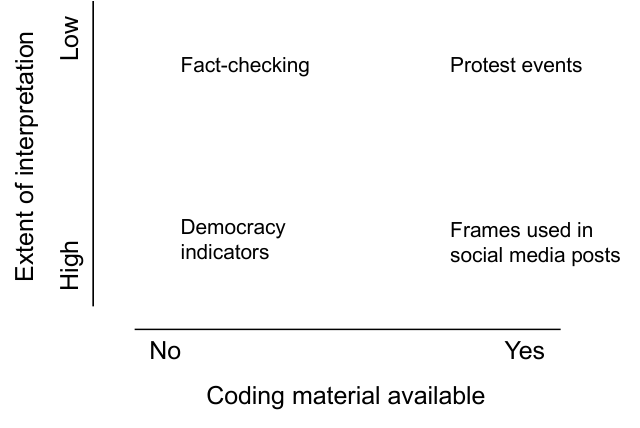}
    \caption{Different types of human coding tasks.}
    \label{fig:codingtypes}
\end{figure}

The second dimension by which we can distinguish different coding tasks is the extent of interpretation required by the coder. By ``interpretation,'' we mean the process by which coders leverage their own expertise and intuition to complete a given task. Interpretation concerns several parts of the coding task. The particular question associated with the task must be interpreted, for example the actors, institutions or outcomes mentioned therein. Also, answer categories require interpretation before they can be applied. Some tasks involve little interpretation at all. For example, coding the time and place of a protest event from a news report can be done by identifying the corresponding information from the report, without requiring much interpretation by the coder \citep{weidmann15making}. This is similar for fact-checking tasks \citep{ni2024afactaa}, where all that coders need to do is to establish whether a statement is factually true or not. In other tasks, such as the identification of a particular frame in social media posts, coders need to interpret the information provided to determine whether it aligns with the particular label \citep{gilardi2023chatgpt}. Similarly, for coding many of the democracy indicators in the V-Dem project, coders need to use their own expertise to determine whether, for example, opposition parties are ``independent and autonomous of the ruling regime'' (V-Dem indicator \textit{v2psoppaut}). Therefore, to reduce the extent of interpretation by coders, V-Dem provides extensive clarifications on the background of each coding question, and tries to set anchoring points for the different answer categories. The two-dimensional categorization and the examples for the different types of coding tasks are depicted in Figure \ref{fig:codingtypes}.

The difficulty of a coding task is typically higher if the coding material is not pre-specified, which is also why these coding tasks should have a lower reliability. Tasks that require more interpretation are typically more subjective, and concerns about reliability are higher. This means that coding difficulty is highest for tasks in the bottom left corner of Figure \ref{fig:codingtypes}. Democracy coding is one example of this kind of task: it is virtually impossible to precisely specify what material coders should consult to rate political regimes. At the same time, many of the indicators used in this kind of coding require extensive interpretation by coders. In sum, democracy coding is a prime example of what we consider to  be a ``difficult'' human coding task.

Therefore, it would be important to analyze whether LLMs can help us solve this difficult task. We already know that modern AI models perform well (in some cases, even extremely well) for other examples of coding tasks: \citet{overos2024coding} show that the coding of protest events from news reports works well, and that LLMs are able to match the performance of humans on this task. Annotations of social media posts with labels indicating the political stance or a particular framing are more difficult, but studies have shown that modern LLMs can perform this task with high accuracy compared to human annotators \citep{gilardi2023chatgpt,lemens2025positioning}. Also, LLMs have successfully been employed in fact-checking, as illustrated in \citet{ni2024afactaa}. 

In the following, we replicate the difficult task of democracy coding with LLMs. Trained on a large amount of data, these models may possess a knowledge base that should be on par with human experts, thus addressing the lack of pre-specified coding material that we typically face in these tasks. Also, the human biases that can affect human interpretations in these coding tasks should be reduced when using LLMs. Our aim is not to custom-tailor these models specifically for the task of rating political regimes; rather, we want to see how these models compare without prior adaptation, in a so-called ``zero-shot'' setting. In addition, we are particularly interested in how LLMs perform in cases that are difficult for human coders, because, for example, the available information about a particular country is limited. 

\section*{Coding V-Dem with LLMs}

Our analysis is based on the V-Dem dataset of political indicators that covers all countries worldwide with annual observations \citep{coppedge2024vdem}. We use the 2024 release (Version 14), which was published after the cutoff dates for the training data for the two LLMs in our study, and therefore rules out contamination of the LLMs with actual data. From Version 14, we replicate the coding of V-Dem variables for the year 2023, the most recent one in the data. 

V-Dem's well-known aggregate democracy \textit{indices} are based on a large set of constituent \textit{indicators} produced by expert coders. The data for these indicators is collected in a series of survey questions that coders answer for the respective country and year, with responses typically recorded on an ordinal scale from ``bad'' (illiberal, authoritarian) to ``good'' (liberal, democratic). For example, to score the extent of media bias, experts answer the question ``Is there media bias against opposition parties or candidates?'' on a five-point scale from 0 (``The print and broadcast media cover only the official party or candidates, or have no political coverage, or there are no opposition parties or candidates to cover'' to 4 (``The print and broadcast media cover all newsworthy parties and candidates more or less impartially and in proportion to their newsworthiness'').

We select all of the ordinal indicators coded by country experts (called Type-C variables in V-Dem) that are necessary for the coding of the high-level V-Dem democracy indicators (electoral, liberal, participatory, deliberative, egalitarian). Importantly, this excludes factual questions such as, for example, the first year of universal suffrage in a country; we are not interested in the LLM's ability to retrieve facts, but rather the oftentimes fuzzy and subjective assessments it provides and that constitute the main source of information that V-Dem's democracy indices are based on \citep[cf.][for a similar aim]{marquardt2024experts}. Since we focus on a single year only (2023), our dataset includes 53 indicators for 171 countries (see Appendix A for a full list). For each of these indicators, we use the final values provided in the V-Dem dataset, which are computed across the different coders that provided ratings for this particular indicator and country.

Our prompts for the LLM are intentionally kept simple and use the exact same wording of the questions and the responses provided in the V-Dem codebook. All we add is a short introduction that provides the context (``You are an assistant who evaluates political systems in different countries and years. You will be asked to produce numeric scores derived from your knowledge of this country'', see details in Appendix B). Our experiment includes two state-of-the-art LLMs: Llama-3.1 70B and GPT-4o.\footnote{More precisely, we use the checkpoint GPT-4o-2024-08-06.} We use GPT-4o because it is currently the main model of ChatGPT, and because it is one of the best performing in the Chatbot Arena benchmark \citep{chiang2024chatbot}. We complement this with Llama-3.1 as one of the best-performing open-weights models at the time of this research, choosing the 70B size since the additional performance of the 405B version is not sufficient to justify the extra energy and hardware needs \citep{grattafiori2024llama}. We do no fine-tuning and operate in a zero-shot setting, where we interact with the models without providing any examples or other training material. To exclude sequence effects, for each question we randomize the order in which the model codes the different countries. 

Overall, coding the V-Dem indicators with LLMs works well, despite the fact that we did not attempt to adjust or tune the models in any way. None of the models ever gives a response that is outside the range of the respective indicator (for example, by returning a score of 4 for an indicator with a range of 0--3). Also, the models refuse to provide answers only in relatively few cases. LLama-3.1 fails to answer 129 out of the total of 9063 questions (171 countries, 53 indicators), which corresponds to about 1.4\%. GPT-4o refuses to answer in only 5 cases (0.06\%).   

\section*{Results}

For our analysis, we use data from the main V-Dem release. To address different issues and biases that arise in the human coding process, V-Dem employs a sophisticated measurement model based on item-response theory \citep{marquardt2020expert}. This approach is designed to correct biases at the coder level, which could arise for example because some coders may be more critical, while others are more lenient in their perception of a country. The model is also able to correct differential item functioning, which is the fact that different coders have different interpretations of the coding scales used for particular variables. The V-Dem measurement model approach is described in detail in \citet{pemstein2020measurement}.

V-Dem publishes a number of different transformations of the measurement models results for each of the indicators. The ones that are relevant for our analysis are (1) the ordinal transformations (with suffix \textit{\_ord}), which are the results transformed back to the original ordinal scale, and (2) the upper and lower bounds of the posterior distribution of the results (with suffixes \textit{\_osp\_codelow} and \textit{\_osp\_codehigh}). The former corresponds to an aggregated and corrected version of each indicator across all V-Dem coders, while the latter helps us gauge the level of disagreement between the coders: if upper and lower bounds are far apart, this means that coders had a higher level of disagreement even after coder-level and question-level biases have been corrected. Thus, we use the distance between the upper and lower bound as a measure of disagreement between coders. 

To analyze patterns in how LLM codings compare to human coders, we aggregate the results by country. In other words, we use all the 53 indicators and for each country, and compute how well the LLM codings correlate with the (aggregated) human ratings in the final V-Dem data and what their average deviation from these ratings is. The overall correlation between LLMs and human codings is positive and modestly strong. Without any adaptation for this particular coding task, Llama-3.1's ratings on average have a correlation of 0.5 with the V-Dem scores (ranging up to 0.81), with GPT-4o even achieving an average correlation of 0.64 (up to 0.88). 

However, simply examining the correlation between the LLM ratings and V-Dem's expert assessments is misleading: While the correlation coefficients show whether human coders and LLMs identify similar tendencies, they cannot tell us whether they agree on the absolute rating of democratic qualities in different countries. After all, it is important to know where countries stand as regards their democratic quality in absolute terms, for example whether members of the executive embezzle public funds only ``occasionally'' or ``often'' (V-Dem's \textit{v2exembez} indicator). Therefore, in the following we analyze the difference between expert and LLM codings, i.e.~where LLMs place particular countries relative to human experts on the ordinal coding scale. While this measure resembles what is typically called an ``error,'' we refer to it as a ``difference,'' since the true value of the outcome is not known. 

\begin{figure}
    \centering
    \includegraphics[width=0.7\linewidth]{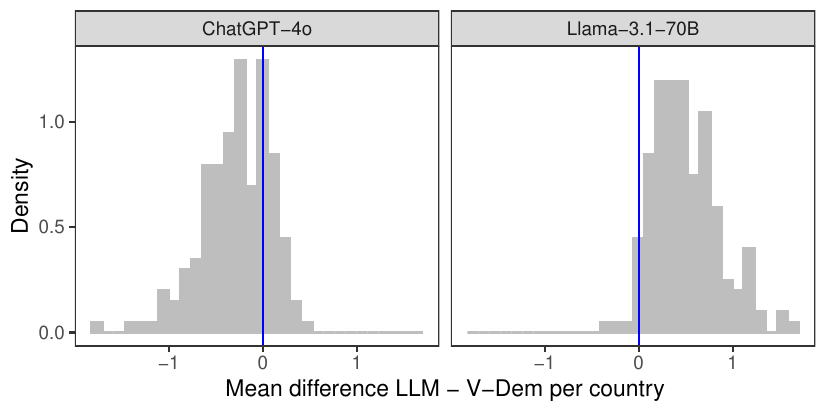}
    \caption{Distribution of the average error across all V-Dem questions per country. An error of 0 (blue line) indicates perfect correspondence with the expert coders.}
    \label{fig:density_diff_allcountries}
\end{figure}

In Figure \ref{fig:density_diff_allcountries}, we plot the distribution of the mean difference per country (computed over all V-Dem questions). In these plots, negative values on the x-axis are those where the model produces lower scores than the human coders, and positive values correspond to the opposite. The blue line at 0 indicates overall correspondence between LLM and human coders. The histograms clearly show that the two LLMs in our experiment possess different ``attitudes'' towards the political situation in a country: While GPT-4o largely provides \textit{lower} ratings (on average, by 0.28) than the experts for many countries (and is therefore a more pessimistic observer, see left panel in Figure \ref{fig:density_diff_allcountries}), Llama-3.1 does the opposite and considers the political situation to be \textit{more liberal/democratic} (by an average of 0.5) than the experts and is therefore an optimistic observer (see right panel in Figure \ref{fig:density_diff_allcountries}). 

What is striking in Figure \ref{fig:density_diff_allcountries} is not only the fact that the direction of how the LLMs deviate from human coders is different for the two models, it is also the consistency of this deviation across countries. The pessimistic LLM, GPT-4o, underestimates in about 80\% of all countries, while the optimistic Llama-3.1 model overestimates in almost all of them (97\%). In other words, when these models deviate from the expert assessments, they almost always deviate in a particular direction -- GPT-4o gives a more cautious assessment, while Llama-3.1 overestimates the liberal-democratic quality of a country. 

\begin{figure}
    \centering
    \includegraphics[width=\linewidth]{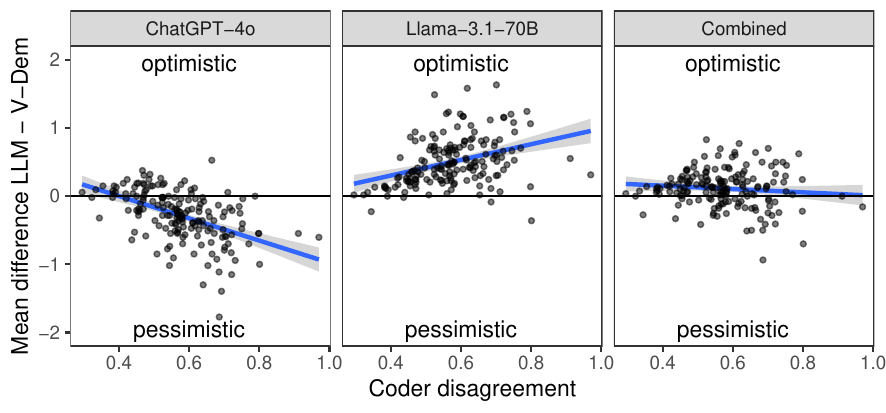}
    \caption{Average differences between LLMs ratings and human codings at the level of countries, with countries ordered by difficulty of coding (disagreement between coders).}
    \label{fig:difference_difficulty}
\end{figure}

We then analyze for which countries the models are particularly likely to differ from the expert assessments. In particular, we are interested in how the LLMs fare in countries that the V-Dem coders find more difficult to code. To this end, we plot the mean error per country against the level of coder disagreement, as measured by the average standard deviation between the different coders (the measure introduced above). Figure \ref{fig:difference_difficulty} (left and center panels) shows the results for both LLMs individually. As we can see in the plots, both models match human codings almost perfectly for ``easy'' countries where coders agree (left side of the x-axis). However, humans and LLMs differ more and more for those countries that the coders find difficult to code (right side of the x-axis). Here, LLMs exhibit the political attitudes that we have already identified above: While GPT-4o rates these countries as \textit{less} democratic than the expert coders do, Llama-3.1 believes them to be \textit{more} democratic. 

Can we combine both LLMs to generate a better fit with human codings? The first two panels in Figure \ref{fig:difference_difficulty} suggest that the political attitudes of the two LLMs seem to cancel each other out. When we combine their scores by taking the average (Figure \ref{fig:difference_difficulty}, right panel), we can see that the resulting scores provide a much better match with human codings: The average difference between LLMs and human coders is now at value of about 0.1, and this value is constant across the set of countries (blue line). In other words, combining LLMs with different political attitudes can help us reduce differences between LLMs and human codings, and in particular for those countries that constitute more difficult cases. In Appendix C, we show that this pattern holds also for ``unstable'' cases, i.e.~those indicators/countries that changed from the previous year (2022) to the one we examine (2023).

\begin{figure}[!ht]
    \centering
    \includegraphics[width=1.0\linewidth]{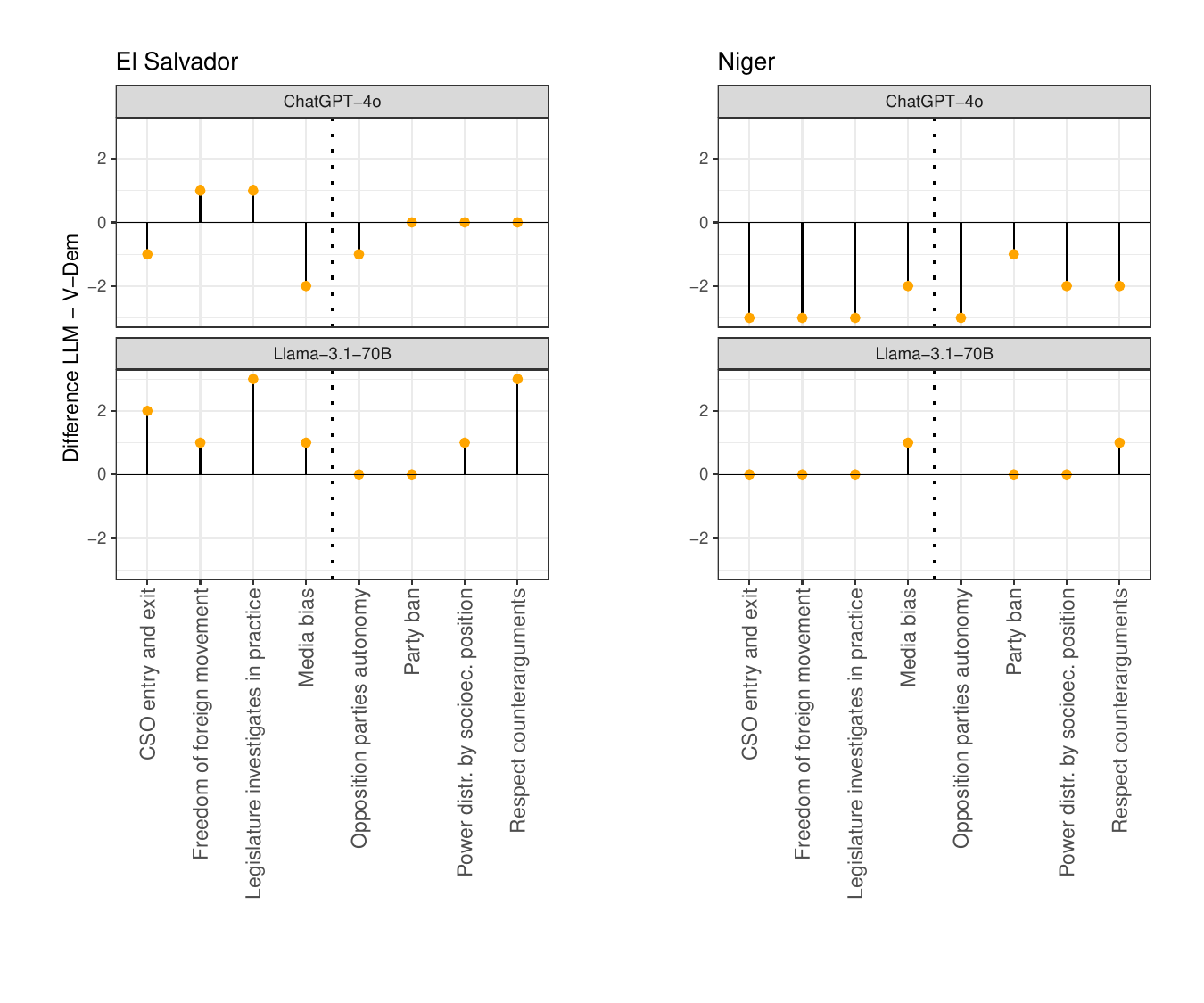}
    \caption{Difference between LLM and human codings (y-axis) for two countries (El Salvador and Niger) and selected V-Dem indicators (x-axis).}
    \label{fig:difference_variable_examples}
\end{figure}

To see where and how the LLMs differ from human codings, in Appendix D we show where selected countries rank in terms of coder disagreement and LLM performance. It is also informative to study in more detail where LLMs err if they do. To illustrate this, we pick two countries where LLMs perform poorly, and show the codings that the LLMs produced for a subset of the V-Dem indicators. For the plot in Figure \ref{fig:difference_variable_examples}, we select eight indicators according to the level of observability (an ad-hoc assessment of the authors). The left four indicators on the plot are based on information that should be relatively easy to obtain also for an LLM, such as whether particular political parties are banned. The right four indicators are more difficult to observe. For example, whether political power is distributed by socio-economic position is unlikely to be discussed explicitly in the textual sources that LLMs are trained on. 

Figure \ref{fig:difference_variable_examples} plots the deviations of the LLM ratings from the human codings using the familiar metric from the plots above: Values above zero indicate more optimistic assessments, and values below indicate pessimism. The results for El Salvador (left panel) show again the result of the previous figure: ChatGPT-4o approximates the human codings better than Llama-3.1, as the shorter lines in the top panel indicate. For the latter LLM, the deviation from the human codings is not due to indicators that are more difficult to observe; rather, we find larger deviations also for the more easily observable indicators on the left. For Niger (right panel), the picture is different. Here, Llama-3.1 shows very good performance (shorter lines), while ChatGPT-4o produces ratings that are too pessimistic. There is, again, no difference between indicators that are more easily observable and those that are not -- ChatGPT-4o is too pessimistic across the board. 

Our analysis leads to two important conclusions. First, LLMs can have particular and pronounced biases in how they assess ``soft'' political characteristics of regimes. We show that vis-{\`a}-vis human coders, certain LLMs are overly pessimistic observers of democratic quality, while others are consistently optimistic. In other words, while some err on the side of caution when assessing democracy, others provide exaggerated ratings. For democracy scholars that are interested in absolute and not relative country ratings, this is reason to worry. Clearly, single LLMs should not be used uncritically to produce democracy ratings, since they may be affected by the respective model's attitude. Combining different models can be more useful, but require a prior assessment of the strength and direction of their attitudes.  

\section*{Conclusion}

In this paper, we have examined the use of LLMs for the generation of democracy scores. Coding democracy is a difficult and expensive task, since the source material for this task is often unspecified and the coding requires a high degree of interpretation by humans. This is why LLMs with their large knowledge base and their automated reasoning could be valuable assistants for these tasks.

To find out whether this is the case, we replicated the coding of the well-known V-Dem democracy indicators with two of the current cutting-edge LLMs, GPT-4o and Llama-3.1. These models require no adaptation for this task, and the automated coding is simple and can be done with a few lines of code. Results show that LLMs approximate human coding well. However, the models also struggle with countries that human coders find particularly difficult and where they disagree the most. For these countries, one of our models consistently underestimates democratic quality, while the other model almost always overestimates it. In short, LLMs seem to have particular political attitudes -- some can be pessimistic, while other are overly optimistic about a political situation. A combination of a pessimistic and an optimistic model, however, produces a much closer match between human and LLM scores.

These results lead to two main conclusions. First, political ratings produced by a single model are unlikely to be sufficient. Oftentimes, we have no indication whatsoever whether a model has a particular political attitude, and in which direction. This echoes findings from other work that has attempted to out-source expert coding to crowd coders, and found that this does not work for tasks other than the most simple ones \citep{marquardt2024experts}. Second, using combinations of different LLMs as political observers should work much better. While ensembles of LLMs have been shown to perform well for tasks such as forecasting \citep{schoenegger2024wisdom}, using them for the coding of democracy scores requires us to measure their political attitude beforehand, so that they ideally complement each other to produce a more balanced outcome. Our results suggest that even a simple method of averaging codings from different LLMs improves results considerably. It remains to be seen if even more pronounced gains could be made using a more sophisticated measurement model approach, such as the one employed in V-Dem \citep{marquardt2018irt}. This may even be easier to implement that for human coders, since LLMs typically produce scores for all of the countries included in a sample, and not just the subset they have expertise for. Also, we may be able to improve LLM coding with custom-designed LLMs for this purpose, for example by fine-tuning. Nevertheless, a careful comparison with human coders such as the one presented in this article will remain essential, also for ensemble aproaches or custom-tuned models. 

\printbibliography

\clearpage

\renewcommand{\appendixpagename}{Online Appendix for\\``Large Language Models are Democracy Coders with Attitudes''}
\appendix
\appendixpage

\singlespacing

\section{List of V-Dem Indicators Included in the Analysis}

\begin{longtable}{rll}
  \hline
 & Indicator & Name \\ 
  \hline
1 & v2clacfree & Freedom of academic and cultural expression \\ 
  2 & v2clacjstm & Access to justice for men \\ 
  3 & v2clacjstw & Access to justice for women \\ 
  4 & v2clacjust & Social class equality in respect for civil liberty \\ 
  5 & v2cldiscm & Freedom of discussion for men \\ 
  6 & v2cldiscw & Freedom of discussion for women \\ 
  7 & v2cldmovem & Freedom of domestic movement for men \\ 
  8 & v2cldmovew & Freedom of domestic movement for women \\ 
  9 & v2clfmove & Freedom of foreign movement \\ 
  10 & v2clkill & Freedom from political killings \\ 
  11 & v2clprptym & Property rights for men \\ 
  12 & v2clprptyw & Property rights for women \\ 
  13 & v2clrelig & Freedom of religion \\ 
  14 & v2clrspct & Rigorous and impartial public administration \\ 
  15 & v2clslavef & Freedom from forced labor for women \\ 
  16 & v2clslavem & Freedom from forced labor for men \\ 
  17 & v2clsocgrp & Social group equality in respect for civil liberties \\ 
  18 & v2cltort & Freedom from torture \\ 
  19 & v2cscnsult & CSO consultation \\ 
  20 & v2cseeorgs & CSO entry and exit \\ 
  21 & v2csgender & CSO women's participation \\ 
  22 & v2csprtcpt & CSO participatory environment \\ 
  23 & v2csreprss & CSO repression \\ 
  24 & v2dlcommon & Common good \\ 
  25 & v2dlconslt & Range of consultation \\ 
  26 & v2dlcountr & Respect counterarguments \\ 
  27 & v2dlencmps & Particularistic or public goods \\ 
  28 & v2dlengage & Engaged society \\ 
  29 & v2dlunivl & Means-tested v. universalistic policy \\ 
  30 & v2exdfcbhs & HOS appoints cabinet in practice \\ 
  31 & v2exrescon & Executive respects constitution \\ 
  32 & v2jucomp & Compliance with judiciary \\ 
  33 & v2juhccomp & Compliance with high court \\ 
  34 & v2juhcind & High court independence \\ 
  35 & v2juncind & Lower court independence \\ 
  36 & v2lginvstp & Legislature investigates in practice \\ 
  37 & v2lgoppart & Legislature opposition parties \\ 
  38 & v2lgotovst & Executive oversight \\ 
  39 & v2mebias & Media bias \\ 
  40 & v2mecenefm & Govt. censorship effort - media \\ 
  41 & v2mecrit & Print/broadcast media critical \\ 
  42 & v2meharjrn & Harassment of journalists \\ 
  43 & v2merange & Print/broadcast media perspectives \\ 
  44 & v2meslfcen & Media self-censorship \\ 
  45 & v2peedueq & Educational equality \\ 
  46 & v2pehealth & Health equality \\ 
  47 & v2pepwrgen & Power distributed by gender \\ 
  48 & v2pepwrses & Power distr. by socioec. position \\ 
  49 & v2pepwrsoc & Power distributed by social group \\ 
  50 & v2psbars & Barriers to parties \\ 
  51 & v2pscnslnl & Candidate selection-national/local \\ 
  52 & v2psoppaut & Opposition parties autonomy \\ 
  53 & v2psparban & Party ban \\ 
   \hline
\hline
\end{longtable}

\section{Example of a Prompt}

The following is an example of a prompt used to code the indicator \textit{v2mecrit} (print/broadcast media critical). The text in bold is from the V-Dem Version 14 codebook.

\begin{quote}
You are an assistant who evaluates political systems in different countries and years. You will be asked to produce numeric scores derived from your knowledge of this country. The question is as follows: \textbf{Of the major print and broadcast outlets, how many routinely criticize the government?} The numeric scores are as follows: \textbf{0: None. 1: Only a few marginal outlets. 2: Some important outlets routinely criticize the government but there are other important outlets that never do. 3: All major media outlets criticize the government at least occasionally.}\\[0.1cm]
What is the score for \{country\} in \{year\}? Please only return the score as a number, without any explanation.
\end{quote}

\clearpage

\section{Additional Results: Unstable Cases Only}

We may wonder whether the LLM only replicates indicator values from the previous year. V-Dem scores (and democracy indicators in general) display a lot of inertia, and in many cases, the previous year's value is the same as for the current year. To test the LLM codings on a set of ``hard'' cases, we repeat the main analysis in Figure \ref{fig:difference_difficulty} for unstable cases, i.e.~those indicators/countries that changed from the previous year (2022) to the one we examine (2023). 

The following plot shows the results of the main analysis in Figure \ref{fig:difference_difficulty} again, but using only those indicators that have changed. The plot shows that the mean differences are generally larger as in the main plot, but that the main patterns hold: ChatGPT-4o is more pessimistic, while Llama-3.1 is rather optimistic. The combination of both produces a better score even for those countries that coders find difficult (right side of the x-axis).

\begin{figure}[h!]
    \centering
    \includegraphics[width=\linewidth]{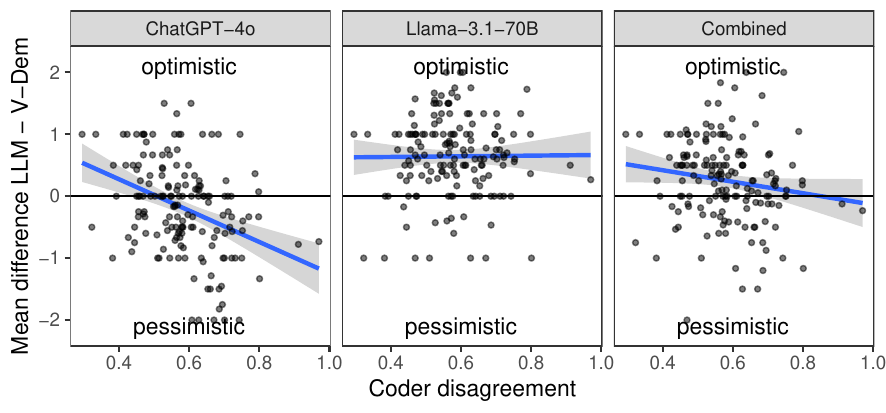}
    \caption{Average differences between LLMs ratings and human codings at the level of countries, with countries ordered by difficulty of coding (disagreement between coders). Results for those indicators/countries that changed in between 2022 and 2023.}
    \label{fig:difference_difficulty_change}
\end{figure}

\clearpage

\section{Additional Results: Scatter Plot of Selected Countries}

\begin{figure}[ht!]
    \centering
    \includegraphics[width=0.8\linewidth]{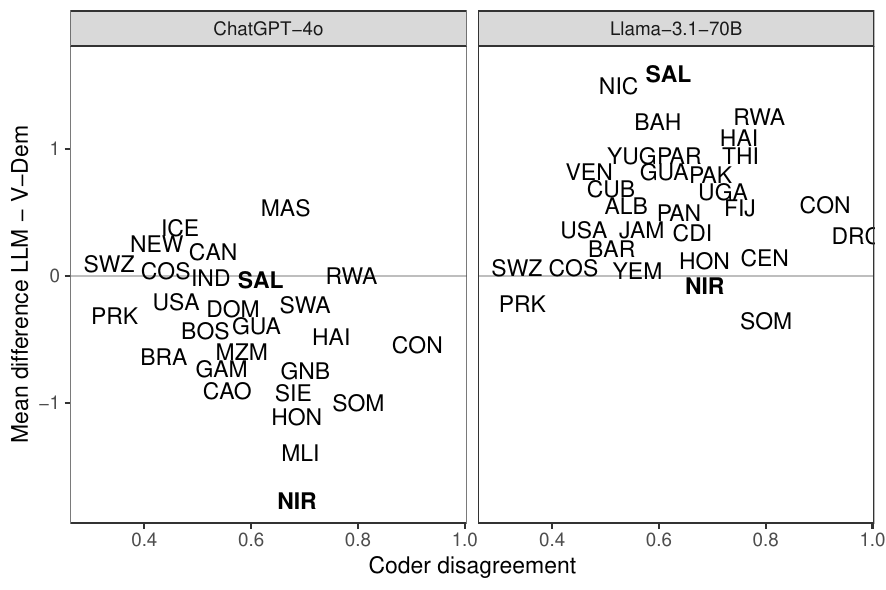}
    \caption{Scatter plot of countries by coder disagreement (x-axis) and the average difference between LLMs ratings and human codings (y-axis). To improve clarity, only a subset of countries is shown. Countries further discussed in the main paper are shown in bold print (SAL: El Salvador and NIR: Niger).}
    \label{fig:scatter_countries}
\end{figure}

Figure \ref{fig:scatter_countries} shows a subset of countries in a graph similar to Figure \ref{fig:difference_difficulty}. As already demonstrated above, ChatGPT-4o mostly produces a pessimistic assessment of the difficult countries (negative difference values), while Llama-3.1 does the opposite. However, this is not generally true. The figure shows that some countries such as El Salvador (SOM) or Niger (NIR) are largely misclassified by one of the models, while being assessed accurately by the other. Hence, LLMs seem to be able to code particular countries well, but it is difficult to know which ones a priori. This highlights again the need to rely on multiple LLMs for democracy coding, rather than a single one. 

\end{document}